# Atomically-thin micas as proton conducting membranes

L. Mogg[1,2], G.-P. Hao[1,3], S. Zhang[1,4], C. Bacaksiz[5], Y. Zou[6], S. J. Haigh[6], F. M. Peeters[5], A. K. Geim[1,2], M. Lozada-Hidalgo[1,2]

[1]National Graphene Institute, The University of Manchester, Manchester M13 9PL, UK
[2]School of Physics and Astronomy, The University of Manchester, Manchester M13 9PL, UK
[3]State Key Laboratory of Fine Chemicals, School of Chemical Engineering, Dalian University of Technology, Dalian 116024, China
[4]Key Laboratory for Green Chemical Technology of Ministry of Education, School of Chemical Engineering and Technology, Tianjin University, Tianjin 300072, China
[5]Departement Fysica, Universiteit Antwerpen, Groenenborgerlaan 171, B-2020 Antwerp, Belgium
[6]School of Materials, The University of Manchester, Manchester M13 9PL, UK

**Monolayers of graphene and hexagonal boron nitride (hBN) are highly permeable to thermal protons[1,2]. For thicker two-dimensional (2D) materials, proton conductivity diminishes exponentially so that, for example, monolayer $MoS_2$ that is just three atoms thick is completely impermeable to protons[1]. This seemed to suggest that only one-atom-thick crystals could be used as proton conducting membranes. Here we show that few-layer micas that are rather thick on the atomic scale become excellent proton conductors if native cations are ion-exchanged for protons. Their areal conductivity exceeds that of graphene and hBN by one-two orders of magnitude. Importantly, ion-exchanged 2D micas exhibit this high conductivity inside the infamous gap for proton-conducting materials[3], which extends from ~100 °C to 500 °C. Areal conductivity of proton-exchanged monolayer micas can reach above 100 S $cm^{-2}$ at 500 °C, well above the current requirements for the industry roadmap[4]. We attribute the fast proton permeation to ~5 Å-wide tubular channels that perforate micas' crystal structure which, after ion exchange, contain only hydroxyl groups inside. Our work indicates that there could be other 2D crystals[5] with similar nm-scale channels, which could help close the materials gap in proton-conducting applications.**

Ion exchangers are non-soluble materials that contain ions within their crystal structure. These ions are easily substituted with other ions of the same polarity, if the material is placed in contact with suitable electrolytes[6]. In essence, ion exchangers act as sponges that can absorb and release ions. Micas are well-known ion exchangers[7–10]. They consist of aluminosilicate layers that are normally covered with cations such as, for example, $K^+$. These native species can be exchanged for other ions such as $H^+$, $Li^+$ or $Cs^+$, which adsorb on the surface of aluminosilicate layers and can also absorb in between them (see Fig. 1). Ion exchange at micas' surfaces proceeds much faster than that within the interlayer space, taking seconds rather than hours[8,10]. It is particularly easy to substitute native ions with protons ($H^+$) as shown by surface force[11–13], XPS[14], X-ray reflectivity[15,16], AFM[17,18], NMR[19] and zeta-potential[20] experiments. Besides being proton exchangers, micas exhibit a relatively sparse crystal structure. Their basal planes contain hexagonal rings of ~5.2 Å in size (Fig. 1b), which are considerably larger than the rings making up, e.g., graphene and $MoS_2$ (~2.5 and 3.2 Å, respectively). From this perspective, micas can be considered as aluminosilicate slabs pierced by tubular channels as shown in Fig. 1a. The channels are not empty but filled with hydroxyl ($OH^-$) groups; which resembles proton-conducting 1D chains in water[21] (Fig. 1a). In this report, we investigate whether these atomic-scale channels in micas allow for proton permeation so that few-layer micas could be used as proton-conducting membranes, despite their relatively large thickness (~10 Å for monolayer micas as compared to 6.5 Å for impermeable $MoS_2$).

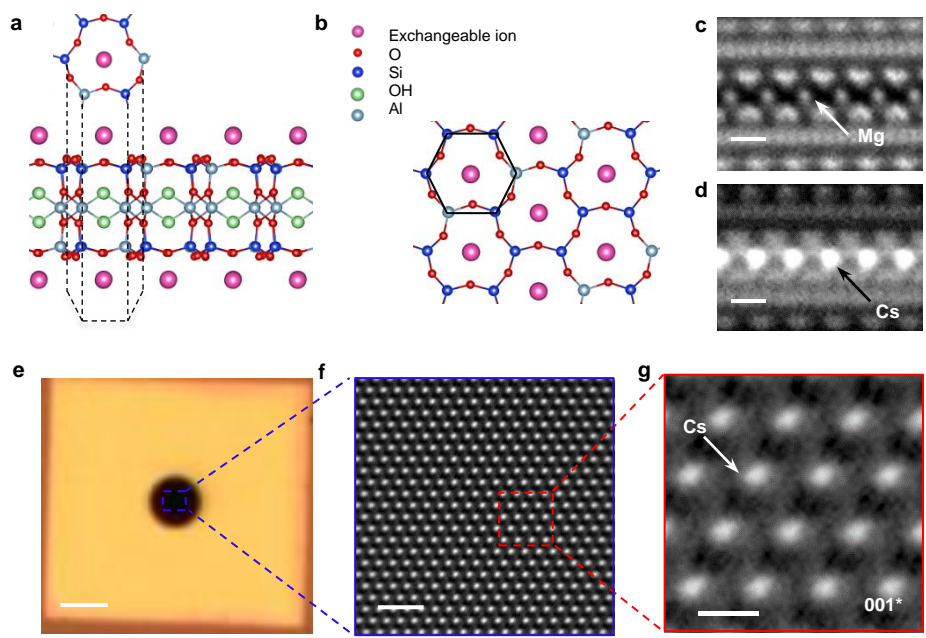

**Figure 1 | STEM characterization of ion-exchanged micas. a**, Cross-sectional schematics of a monolayer mica. The crystal consists of Si-O tetrahedra and Al-O octahedra that form a 2D sheet which we refer to as the aluminosilicate layer. Exchangeable ions (magenta balls) are adsorbed onto these layers. **b**, Corresponding plan-view schematic. Si (dark blue), Al (light blue) and O (red) atoms form ~5 Å hexagonal rings (black lines) so that micas can be viewed as stacked aluminosilicate slabs pierced by tubular channels (see the 3D projection indicated by the dashed lines in panel a). The hexagonal channels contain OH groups (green) and exchangeable ions in the center. **c**, Cross-sectional STEM image of a non-exchanged vermiculite using the high-angle annular dark-field (HAADF) mode. The arrow indicates native Mg ions. Other ions comprising the aluminosilicate layers are also visible in the micrograph. Scale bar, 0.5 nm. **d**, Similar HAADF-STEM image but for Cs-exchanged vermiculite. Large bright spots are Cs atoms. Scale bar, 0.5 nm. **e**, Optical micrograph of a bilayer vermiculite device for plan-view STEM imaging. The silicon nitride membrane is seen in yellow and has a hole in the middle over which the bilayer mica is suspended. Scale bar, 2 μm. **f**, HAABF-STEM image of the bilayer device after Cs-ion exchange. Scale bar, 5 nm. **g**, Zoom-in from panel f. Scale bar, 0.5 nm.

Two types of micas (muscovite and vermiculite) were used in our studies. Atomically-thin crystals of these micas we prepared by mechanical exfoliation (Fig. S1)[22] and first investigated using scanning transmission electron microscopy (STEM). To this end, a selected crystal was immersed in a 0.1 M $CsNO_3$ solution at 80 °C for typically a week so that there was sufficient time for Cs-ion exchange through their entire volume. Cs was chosen for this study as a heavy ion that provides bright contrast in STEM. Using the standard mechanical-polishing approach we then prepared thin slices of our micas for cross-sectional STEM imaging[8]. Figs. 1c and d show examples of such atomic-resolution images for natural and Cs-exchanged vermiculite, respectively. Cs atoms in the ion-exchanged samples are easily distinguished from Mg ions (native for vermiculite) because the former give rise to notably brighter spots. For plan-view imaging, we used our thinnest mica crystals (mono- and bi-layers) which after exfoliation were suspended over microfabricated holes (2 μm in diameter) in a silicon nitride membrane[1] (Fig. 1e, Fig. S2). The structures were then immersed in heated $CsNO_3$ for about an hour, which was found to be sufficient for the STEM observation of Cs exchange. Figs. 1f and g show images for one of our bilayer devices. Cs ions form a triangular lattice, in agreement with our density functional theory (DFT) simulations (Fig. S3). Such clear images of micas' basal plane were conspicuously absent in the literature (probably because of the insulating nature of micas) and were possible in our experiments only if we used Cs-exchanged few-layer crystals. Because the plan-view samples were only exposed to the electrolyte for a short time, we believe that the observed Cs ions were adsorbed on the cleaved surface and not necessarily intercalated the interlayer space. Furthermore, imaging of the suspended mica structures allowed us to confirm that ion exchange (using either $Cs^+$ or $H^+$) did not introduce defects into the crystal lattice (Fig. S4), in agreement with previous studies on proton-exchanged[14–18] and ion-exchanged[10,14,18,23] micas.

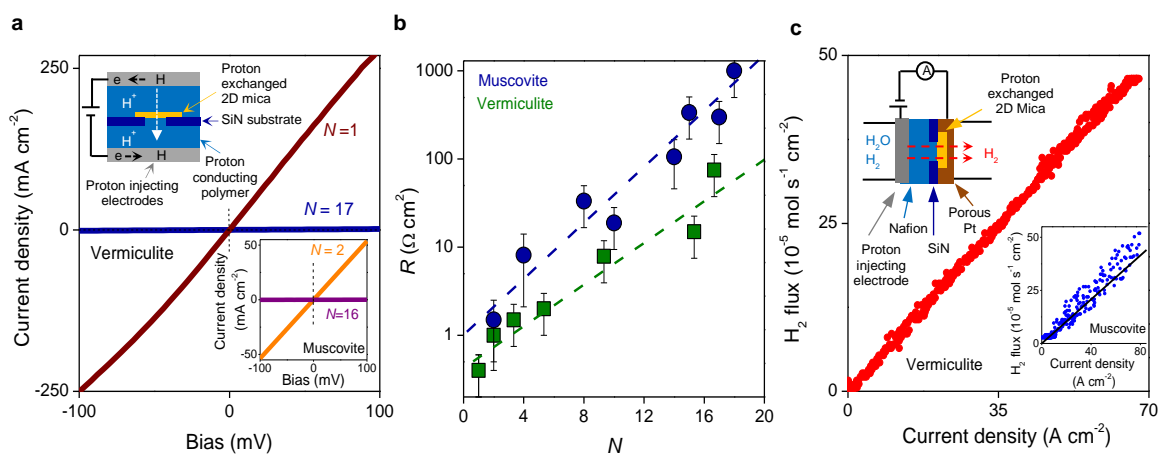

**Figure 2 | Proton transport through 2D micas studied using Nafion-coated devices**. **a**, Examples of *I-V* characteristics for atomically-thin vermiculite (main panel) and muscovite (bottom inset). Top inset, devices' schematic. **b**, Areal resistivity *R* as a function of *N* for both micas. Dotted lines, best exponential fits to the data. **c**, Mass spectrometry of the hydrogen flow through monolayer vermiculite (main panel) and bilayer muscovite (bottom inset). The measured current was induced by negative *V* (typically, < 3 V) applied as shown schematically in the top inset. The electrically driven protons pass through the micas and form hydrogen molecules in the porous Pt electrode: $2H^+ + 2e^- \rightarrow H_2$. The black line in the bottom inset denotes the 100% Faraday efficiency for charge-to-mass conversion.

The possibility of proton transport across the basal plane of atomically-thin micas was studied using electrical measurements and mass spectrometry[1,2]. In these experiments, monocrystals of muscovite and vermiculite were again suspended over micrometer holes similar to Fig. 1e. Then the structure was coated from both sides with a proton-conducting, electron-insulating polymer (Nafion[24]) and proton-injecting electrodes were attached as described previously[1,2] (inset of Fig. 2a and Fig. S2). It is important to note that Nafion is a proton electrolyte and, hence, micas in these devices inevitably become proton-exchanged, regardless of whether or not ion exchange procedures were implemented prior to the Nafion deposition. Nonetheless, we normally performed such initial proton exchange, which involved immersing the suspended devices in a 10 mM acetic acid solution at 80 °C for 1 hour and then thoroughly rinsing them with deionized water. For electrical measurements, the devices were placed in a humid $H_2$ atmosphere to ensure the highest possible proton conductivity of Nafion[1]. The measured current density *I* varied linearly with applied voltage *V* for all the studied devices (Fig. 2a), which allowed us to determine their areal resistivity, *R=V/I*. For both micas, *R* was found to increase exponentially with their thickness (Fig. 2b) so that $R = R_0$

exp($\alpha N$) where $N$ is the number of layers, $\alpha$ is a material-specific coefficient and $R_0 \approx 1$ Ω cm$^2$ is the areal resistivity in the monolayer limit. Note that $R_0$ is much smaller than the corresponding values of ~100 and 10 Ω cm$^2$ for monolayers of graphene and boron nitride, respectively[1]. Such high transparency of micas with respect to protons can be attributed to sparse electron clouds that are formed by the aluminosilicate layers[1,2] (see below).

To corroborate the results of our electrical measurements, the proton flux was also measured directly, using mass spectrometry[1,2]. In this case, one of the Nafion layers was effectively removed and, instead, a porous Pt electrode (~50 nm thick) was sputtered directly onto mica (inset of Fig. 2c, and Fig. S2). The device was then used to separate two chambers: one containing $H_2$ at 100% humidity and the other evacuated and connected to a mass spectrometer. If a positive or zero voltage was applied to the Pt electrode, no gas permeation could be discerned by the spectrometer. This provides a clear proof (in addition to the standard control experiments[1,2]) that the studied mica membranes were impermeable to gases, ruling out any structural defects, in agreement with our STEM analysis. On the other hand, if a negative bias was applied, we readily detected a relatively large $H_2$ flow (Fig. 2c). Crucially, for every two electrons passing through the electrical circuit shown in the inset of Fig. 2c, one hydrogen molecule was detected in the vacuum chamber. This corresponds to 100% efficiency for charge-to-mass conservation as described by Faraday's law of electrolysis: $\Phi = I/2F$, where $\Phi$ is the hydrogen flux, $F$ is Faraday's constant and the factor of 2 accounts for the two protons required to form a hydrogen molecule[1,2].

The mica membranes described above were proton-exchanged. To find out if the exchange increased or decreased their proton transparency, we compared proton-exchanged and natural (non-exchanged) mica devices. They were fabricated in a similar manner but, instead of using Nafion, porous Pt films were sputtered directly onto both sides of mica (inset of Fig. 3a and Fig. S2). If placed in a humid $H_2$ atmosphere, such Pt films are known to serve as source and sink reservoirs for protons[25]. Fig. 3a shows that proton exchange had a strong effect on proton permeation such that the areal conductivity, $G=1/R$, of H-exchanged devices ($G_x$) was ≳100 times larger than $G$ of non-exchanged ones ($G_n$): $G_x \gtrsim 100\, G_n$. Furthermore, the strong dependence of $G_x$ and $G_n$ on both $H_2$ pressure and humidity proved that the observed conductance in both exchanged and non-exchanged micas was provided by protons rather than electron transport (Fig. S5). This also agrees with the fact that, for a given type of mica and its thickness, $G_x$ closely matched the values of $G$ found for Nafion-coated devices (Fig. 2b). These experiments show that protons can transport through both exchanged and non-exchanged micas but proton exchange increases $G$ strongly, by over two orders of magnitude.

Unlike Nafion, the Pt-coated devices could sustain high temperatures ($T$) and, therefore, allowed us to study $T$ dependence of proton transport (Fig. 3a). To ensure that changes in devices' hydration at elevated temperatures did not introduce artefacts into the measured dependence, our devices were normally placed in dry hydrogen. Nonetheless, we also checked in several cases that a humid hydrogen atmosphere gave rise to similar $G_x(T)$. We found that $G_x$ strongly increased with $T$, by a factor of ~600 between 30 °C and 350 °C (Fig. 3b, red curve). The Arrhenius plot reveals two distinct $T$ regions separated by the transition temperature $T_p$ of ~200 °C. In both regions, $G_x$ displayed activated behavior $G_x \propto \exp(-E/kT)$; but with different energy barriers $E$ that we refer to as $E_p$ and $E_e$ below and above $T_p$, respectively ($k$ is the Boltzmann constant). $E_p$ was found to be ~0.2 eV for both muscovite and vermiculite and, within our experimental accuracy of 50 meV, the micas' thickness had little effect on it (Fig. 3b, bottom inset). On the other hand, the activation curves were notably steeper above $T_p$, indicating higher activation energies, $E_e \approx 0.8$ eV and $\approx 0.5$ eV for muscovite and vermiculite, respectively (Fig. 3b and Fig. S6). The Arrhenius plots were reproducible between different devices and displayed no hysteresis during heating and cooling cycles (Fig. S6).

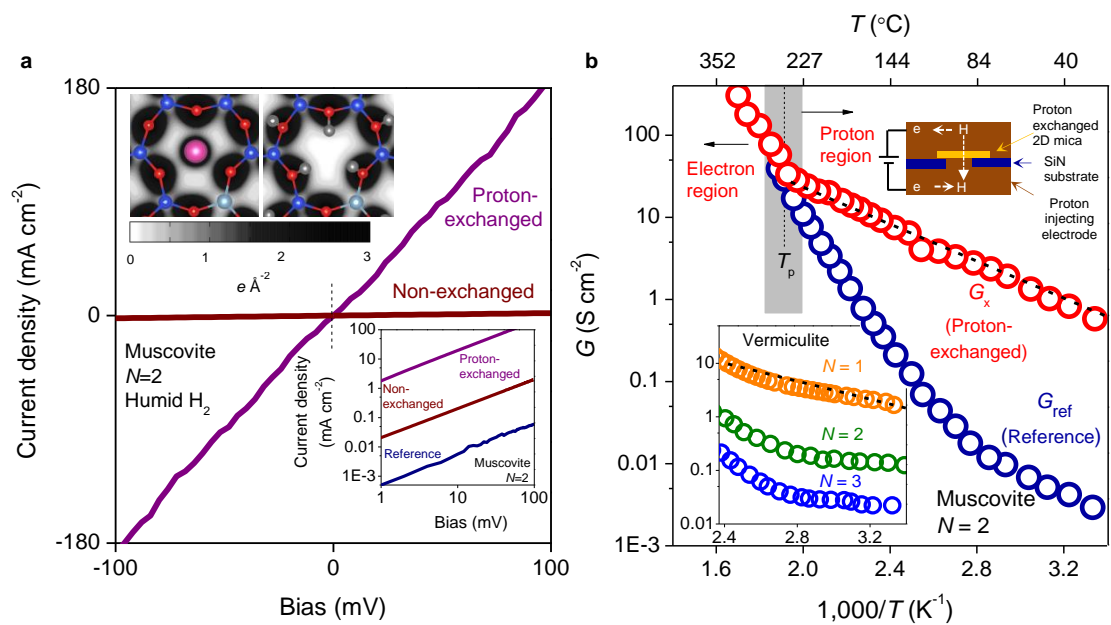

**Figure 3 | Proton transport through micas measured using Pt-coated devices. a**, *I-V* characteristics of proton-exchanged and non-exchanged bilayer mica in humid $H_2$. Top inset, plan view of electron charge density for non-exchanged (left) and proton-exchanged (right) monolayers of muscovite. For clarity, the crystal structure is overlaid as red balls indicating oxygen atoms; dark blue, silicon; light blue, aluminum; grey, hydrogen from proton exchange; magenta, exchangeable ion (potassium in this case). Grey scale, areal density in units of *e* per Å$^2$. Bottom inset: *I-V* characteristics from the main panel (purple and brown) are replotted on the log-log scale to compare them with the *I-V* response of reference devices (blue). **b**, $G_x$ (red) and $G_{ref}$ (blue) measured for bilayer muscovite using porous-Pt contacts as shown schematically in the top inset. The grey rectangle indicates $T_p \approx 250°C$ separating regimes with different activation behavior. Below $T_p$, proton transport dominates $G_x$ whereas electrons become responsible for conductivity at higher *T*. Bottom inset: Similar Arrhenius plots for proton-exchanged vermiculite of different thickness in the *T* range where proton transport is dominant. They show little dependence of $G_x(T)$ on *N*. The black dotted lines indicate $E_p = 0.2$ eV.

The above measurements suggest that two transport mechanisms with different activation energies contribute to $G_x(T)$. To gain further information about them, we studied Pt-coated but non-exchanged devices that were placed in vacuum (referred to as reference devices). In the absence of a source of protons, there is no possibility of proton conductance and only electrons can tunnel through atomically-thin micas leading to finite leakage currents. The bottom inset of Fig. 3a shows that at room *T* the conductivity $G_{ref}$ of our reference devices was ~1,000 times lower than that for similar but proton-exchanged mica. On the other hand, $G_{ref}$ increased with *T* more rapidly than $G_x$, exhibiting activation energy $E_{ref}$ close to $E_e$. Above $T_p$, both reference and proton-exchanged devices displayed essentially the same conductivity, that is, $G_{ref} \approx G_x$ (Fig. 3b). These observations show that the dominant contribution to the micas' conductance above $T_p$ is electron transport. This is not surprising as micas are known to allow activated hopping of electrons between impurities, usually Fe atoms[26]. At lower *T*, proton transport clearly dominates, in agreement with our results using Nafion-coated devices (note that these latter devices did not allow for electron conductance because Nafion films are highly insulating for electrons)[1,2]. Combining the observed $T$ and $N$ dependences (including those in the bottom inset of Fig. 3b), we obtain the following empirical formula for proton transport through micas' basal planes

$$R_p = 1/G_p \approx R_0 \cdot \exp(\alpha N) \cdot \exp(E_p/kT) \tag{1}$$

where $E_p \approx 0.2$ eV and $\alpha \approx 0.3$–0.35 for both studied micas.

The reason for easy proton transport through micas can be qualitatively understood by considering their crystal structure. As shown in Fig. 1a, micas are essentially aluminosilicate slabs perforated by ~5 Å-wide tubular channels. The presence of such channels is substantiated by the DFT calculations in Fig. 3a, which show the areal electron density seen by protons as they try to pierce monolayer mica. If micas are non-exchanged, native ions block the channels' entrances, as evidenced by the dense electron cloud in the center of the hexagonal ring in the inset of Fig. 3a (left panel). After the ions are removed by proton exchange (right panel), clear voids emerge with a size notably larger than that found in similar analyses for graphene and boron nitride monolayers[1] (see also Fig. S7). The remaining electron density inside the tubules is mostly due to two $OH^-$ groups (see Fig. 1a). Being negatively charged, $OH^-$ groups can trap protons translocating along the hexagonal channels. The exponential dependence of $R$ on $N$ (Fig. 2b) can then be attributed to hopping between such proton traps stationed in consecutive layers. Note that it is protons ($H^+$) rather than hydronium ions ($H_3O^+$) that hop across the lattice, since water molecules are sterically excluded[11,12,17].

Our DFT calculations support the above interpretation. They show that the minimum energy path for protons involves the two $OH^-$ groups. These create local energy minima around them (Fig. S8), thus providing the traps between which protons jump. The calculations also yield a maximum energy barrier for proton transport of ~0.35 eV. This agrees well with the experimentally measured $E_p \approx 0.2$ eV, especially if the role of surface-adsorbed water is taken into account. Indeed, protons in surface-adsorbed water are strongly attracted to $OH^-$ groups inside the mica tubules[17]. The first link in the transport path can therefore be expected to be a proton ($H^+$) jump from surface adsorbed $H_3O^+$ towards $OH^-$ groups. Quantum effects during similar jumps were shown to reduce proton transport barriers in graphene and hBN[2] and a similar reduction could be expected in micas.

To conclude, we have shown that proton-exchanged few-layer micas are good proton conductors. In comparison with much thinner monolayers of graphene, hBN and especially $MoS_2$, it is the presence of tubular channels in micas that enables their high proton conductivity. From this perspective, other 2D materials that may be relatively thick but possess similar tubular structures (for example, complex cuprates or 2D metal-organic frameworks) could be good proton conductors or turned into such via proton exchange. In terms of applications, the present work also merits attention. There is a lack of proton conducting materials that can operate between 100 °C and 500 °C, which is referred to as the materials gap[3]. Bridging this gap would enable higher efficiency and lower cost in a large number of energy conversion technologies, which warrants the current intense activity in the field[27,28]. Atomically-thin micas are good candidates to help fill this gap. Indeed, even at moderate temperatures of ~150 °C and in dry hydrogen, their $G_p$ exceeds ~10 S cm$^{-2}$ (Fig. 3b), twice higher than the industry benchmark placed by Nafion 117[29]. Furthermore, micas are highly stable in both oxidizing and reducing atmospheres up to very high $T$ (above 800 °C)[30]. Extrapolating our data to the other side of the materials gap yields extremely high conductance > 100 S cm$^{-2}$ at 500 °C, which may be achievable by growing micas without impurities (such as Fe) to prevent the high-*T* electron leakage characteristic to natural micas. We envision that the reported proton-conducting membranes can be scaled up using, for example, vapor deposition of atomically-thin micas[31] and their subsequent transfer onto a suitable porous substrate, like in the industrial process successfully developed for graphene[32]. Ultrathin mica laminates similar to those reported in ref.[33] can be another viable option for scaling up.

# Atomically-thin micas as proton conducting membranes

L. Mogg, G.-P. Hao, S. Zhang, C. Bacaksiz, Y. Zou, S. J. Haigh, F. M. Peeters, A. K. Geim, M. Lozada-Hidalgo

**Crystal characterization**

Micas consist of two sheets of Si-O tetrahedral groups that sandwich an Al-O octahedral group sheet to form a 2D aluminosilicate layer. Micas are classified as trioctahedral and dioctahedral. In the former, all octahedral groups have a positive ion inside (e.g. $Fe^{2+}$); in the latter, two out of three octahedral groups in the 2D aluminosilicate layer contain a positive ion (e.g. $Al^{3+}$), with the remaining octahedral group lacking a central ion. These 2:1 layers stack on top of each other, separated by interlayer cations[1,2] (e.g. $K^+$) which can be exchanged, to form the crystal. These cations neutralize the excess negative charge in the 2:1 layers (see ref. [1,2] for details of the mica structure).

The micas studied in this work were natural vermiculite, sourced from Lanark, Ontario (Canada) and ruby muscovite, purchased from Agar Scientific. The chemical composition of laboratory grade ruby muscovite is readily found from the suppliers, water (2.99%), $K_2O$ (9.87%), $MgO$ (0.38%), $Fe_2O_3$ (2.48%), Si-O + Al-O (78.67%), $Na_2O$ (0.62%), C (0.44%) and traces. Natural vermiculite was characterized in our experiments to establish its elemental composition. To that end we performed two sets of measurements. First, thermo-gravimetry was used to measure the amount of adsorbed water. Second, inductively coupled plasma atomic emission spectroscopy (ICP-AES) and X-ray spectroscopy were used to determine the elemental composition. To prepare samples for both experiments, bulk vermiculite crystals were first thermally expanded and then ground into powders with grain-size of 61-74 $\mu$m (200-250 mesh).

The weight fraction of adsorbed water was determined by thermo-gravimetry. The vermiculite powders (200 mg) were first dried at 80 $^o$C for 24 h and then weighed. This yielded a 4.9% (9.8 mg) weight loss. Then the sample was further dried at 250 $^o$C for another 24 h. This resulted in a 14.3% (28.6 mg) weight loss. Following the literature, the first weight loss at 80 $^o$C is attributed to the surface adsorbed water whereas the 14.3% loss at 250 $^o$C to the interlayer water[3,4]. This yields a total water mass fraction of 19.2%. The thermo-gravimetry behavior of our samples was consistent with that of typical vermiculites[3,4].

We then determined the elemental composition of our vermiculite via ICP-AES and X-ray spectroscopy. The interlayer cation composition was determined via ICP-AES. To that end, we extracted the cations via ion exchange. Vermiculite powders (400 mg) were dispersed in 1.0 M LiCl (100 mL) and the dispersion was stirred at 80 $^o$C for 24 h. This served to exchange native cations for Li. The suspension was then centrifuged to yield precipitated vermiculite particles and the supernatant (i.e. the dissolved species). This procedure was repeated three times on the precipitate to ensure all interlayer cations were exchanged. The collected supernatants were combined, diluted with 2% $HNO_3$, filtered through a nylon membrane (0.20 $\mu$m pores) and analyzed by ICP-AES. The detected interlayer metal cation mass fractions were calculated to be 2.7% $K^+$ ($K_2O$: 3.26%), 6.45% $Mg^{2+}$ (MgO: 10.75%), 3.01% $Al^{3+}$ ($Al_2O_3$: 5.68%), 1.97% $Fe^{3+}$ ($Fe_2O_3$: 2.81%). This is consistent with the composition of typical vermiculites[5]. After this, we turned our attention to determining the composition of the aluminosilicate slabs. To that end, the precipitate from the previous measurement was washed three times by re-dispersing it in DI water and then dried at 80 °C for 24 h. The collected undissolved powders were weighed and analyzed via energy dispersive X-ray spectroscopy. This revealed that the powders were composed of Si, Al and O, indicating that they are the Si-O/Al-O layers. They were then weighed and found to represent 54.9% mass fraction – again consistent with that of the typical vermiculite[6]. Combined, the thermo-gravimetry and ICP-AES analysis account for 96.6 wt% of the vermiculite. The additional, 3.4% is attributed to ignition and other type of losses during processing as well as possible traces of other elements.

**Device fabrication**

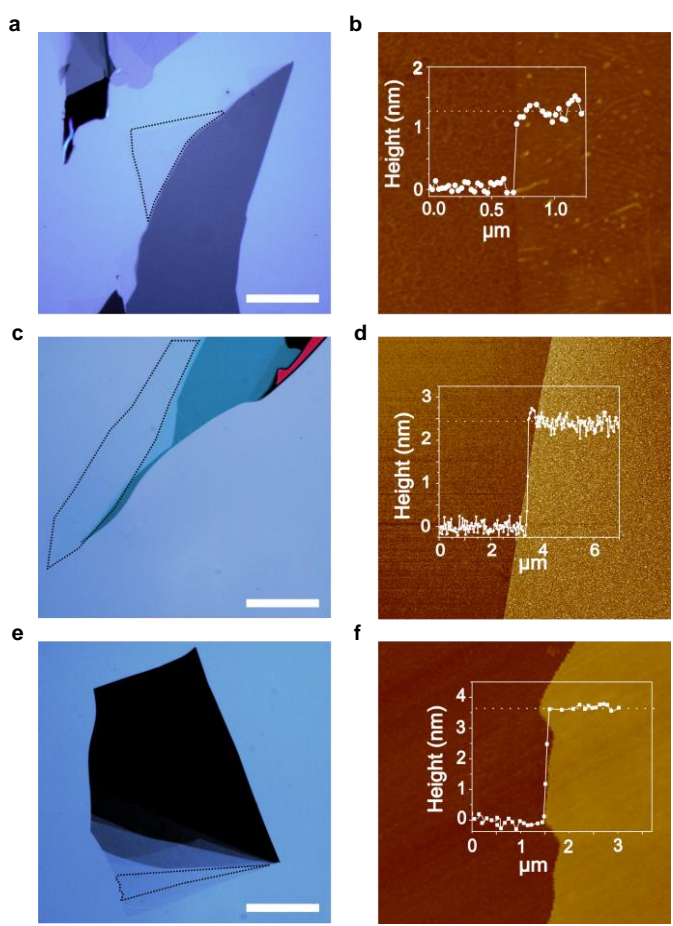

**Supplementary Figure 1 | Flake preparation**. Left panels show optical images of mica flakes. The areas marked with dark lines are mono- (a), bi- (c) and tri-(e) layers. Scale bars, 25 μm. Right panels show AFM images of the crystal edge for the images to the left. Insets in each panel show the height profile of the crystal edge, demonstrating that the crystal are 1-, 2- and 3-unit-cell thick.

The device fabrication process starts with micromechanical exfoliation of the crystals using the dry-transfer technique[7]. Micas can be exfoliated into flakes with precisely controlled thickness. Supplementary Fig. 1 shows optical and AFM images of three mica crystals that are 1, 2 and 3 unit-cell thick. The number of aluminosilicate monolayers in the crystals were unambiguously found using

AFM images and comparing the height the crystal edge ($h$) with the size of the mica unit cell ($c \approx 10$ Å)[8]. Our analysis also took into account the presence of adsorbed species. Large-area monolayer flakes could be isolated from vermiculite. On the other hand, for muscovite, bilayer crystals were the thinnest that could be isolated, which is consistent with stronger interlayer bonding for muscovite[9].

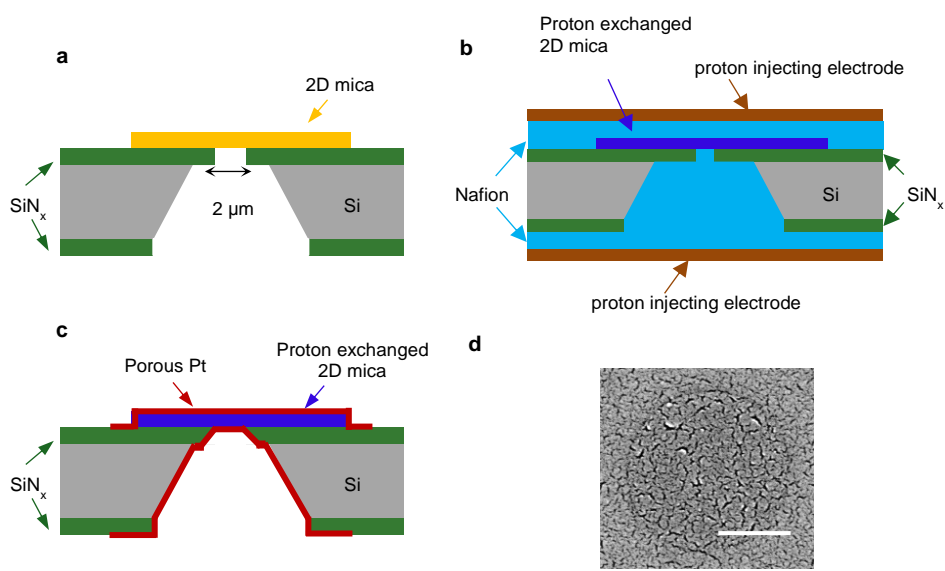

**Supplementary Figure 2 | Device fabrication**. **a**, Schematic of suspended mica devices. **b**, Schematic of Nafion-coated devices. **c**, Schematic of Pt-coated devices. **d**, Electron micrograph of one of our Pt-coated mica devices, top view. The darker circle discernible in the image corresponds to the aperture in the SiN$_x$ substrate over which the mica crystal was suspended. Scale bar, 1 μm.

The next step in the fabrication process was to suspend the exfoliated atomically-thin crystals over microfabricated holes etched in free-standing silicon nitride membranes (Supplementary Fig. 2). To this end, we followed the recipe described in the previous report[10]. The devices were then proton-exchanged. They were immersed for ~1 hour in a heated (~80 °C) 10 mM acetic acid solution, thoroughly rinsed with deionized water, left to dry in air and then heated at ~150 °C to remove remnant moisture. If working with non-exchanged devices, we skipped this step. From this point, the fabrication process differed for Nafion-coated and Pt-coated devices. Nafion-coated devices were fabricated by drop-casting a Nafion solution (5% Nafion; 1100 EW) on both sides of the suspended mica membrane and electrically contacting them with proton-injecting electrodes (PdH$_x$ foil) – see Supplementary Fig. 2b. The whole assembly was then annealed in a humid atmosphere at 130 °C to crosslink the polymer. For Pt-coated devices, instead of coating them with Nafion, thin Pt films (~50 nm) were sputtered directly on both sides of the suspended mica membrane. By controlling the parameters of the sputtering process we obtained porous Pt films (Supplementary Fig. 2d).

Note the similarity of the proton-exchange process with the process involved in fabrication of Nafion-coated devices. In both cases, a heated proton-exchange electrolyte (either acetic acid or Nafion) is in direct contact with the suspended mica membrane. Because of this, all Nafion-coated devices became proton exchanged. To verify this, we fabricated suspended mica devices both proton-exchanged (with acetic acid) and non-exchanged, as described above. These were assembled into Nafion-coated devices. The measured resistance of both devices was the same, which shows that both were proton-exchanged.

Devices for mass spectrometry were fabricated in essentially the same way as Nafion-coated devices. The mica was suspended over holes etched in $SiN_x$ and proton exchanged. One side of the device was coated with Nafion and a proton injecting electrode as described above. However, the opposite side of the devices was not coated with Nafion but instead a porous Pt film was deposited on top. This Pt layer was required to close the electrical circuit and allowed $H_2$ to permeate through the porous film. See ref. [11] for further details.

**Electrical and mass spectrometry measurements**

For electrical measurements, the assembled devices were placed in a chamber with a controlled atmosphere of either $H_2$ at 100% $H_2O$ relative humidity or, alternatively, ~10% humidity (see below). The *I-V* characteristics were measured with a *Keithley* SourceMeter 2636A at voltages typically varying between ±200 mV and using sweep rates <0.1 V min$^{-1}$. The resulting *I-V* characteristics were linear, which allowed us to extract the device resistance. It is important to notice that reporting our data in units of areal resistivity ($\Omega$ cm$^2$) rather than bulk resistivity ($\Omega$ cm) is central to avoid inconsistencies. The very definition of bulk resistivity ($\Omega$ cm) implies that the resistance should scale linearly with the thickness and, hence, cannot be applied to materials like micas, whose resistance scales exponentially with the number of crystal unit cells (*N*). The exponential scaling of proton transport was also observed for other 2D crystals[10].

In control experiments, we measured the resistance of devices with Nafion layers as described above but without a mica membrane. The areal resistivity of these devices was $R$~$10^{-3}$ $\Omega$ cm$^2$, which is ~100 times smaller than that of our most conductive mica devices. This ensured that the electrical response was dominated by the mica membrane and that the Nafion and proton injecting electrode layers added negligibly little into the series resistance. In another set of control measurements, we checked that there was no leakage along the interface between mica and the $SiN_x$ substrate. To rule this possibility out we measured devices where an SU-8 polymer clamp was used to seal the edges of the mica crystal. The resistance of such devices was found to be the same as those without SU-8 clamps, which shows that there was no interfacial leakage. Note that the found exponential dependence on the crystal thickness also rules out any measurable contribution of leakage currents.

For mass spectrometry measurements, devices were clamped with O-rings to separate two chambers: one filled with a gas mixture (10% $H_2$ in Ar, 100% humidity) and the other evacuated and connected to a mass spectrometer. The porous Pt layer faced the vacuum chamber whereas the Nafion layer faced the gas chamber. A dc voltage was applied across the mass-spectrometry devices, and both electrical current and hydrogen flow were measured simultaneously. The mass spectrometer used was an Inficon UL200. As a reference, we measured similar devices prepared as described above but without a mica membrane. For a given voltage bias, both proton and $H_2$ flows were much larger than those for the mica devices. However, the ratio of the $H_2$ flux to the proton current remained the same given by Faraday's law of electrolysis, as dictated by 100% charge-to-mass conservation (see ref. [11] for further details).

**STEM measurements**

Ion exchange in micas was studied using scanning transmission electron microscopy (STEM). Cs was chosen as the exchange ion because it is easily imaged by STEM. Note that ion exchange takes place both in the interlayer space and on the cleaved basal surfaces of mica. To prepare specimens to study interlayer Cs-exchange, we mechanically exfoliated micas on a Si substrate. Two of such substrates were then pressed against each other such that a mica crystal was sandwiched in between them. The Si-mica-Si structure was bound together with epoxy and then mechanically polished using an automatic mechanical polishing instrument (Allied MultiPrep). To perform the Cs exchange, the specimens were immersed in a 0.1 mol $L^{-1}$ $CsNO_3$ for a week and then rinsed in ethanol. On the other hand, to prepare specimens to study surface Cs-exchange, mechanically exfoliated micas were suspended over a hole etched in a free standing $SiN_x$ membrane. The devices were immersed in the same $CsNO_3$ solution for about an hour to perform the Cs-exchange.

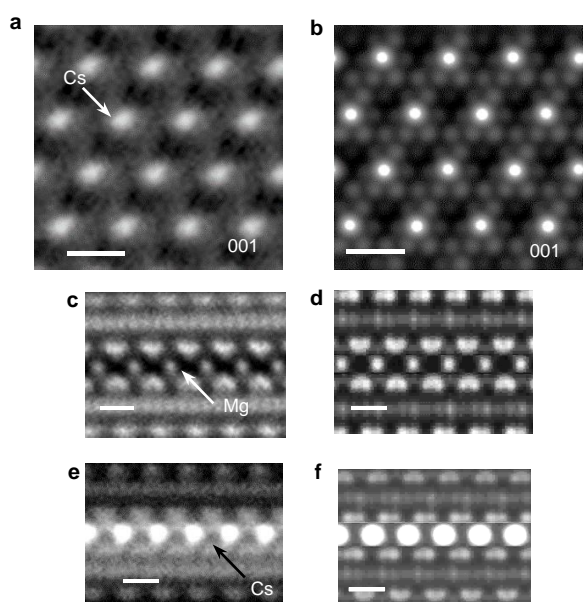

**Supplementary Figure 3 | Ion lattices as seen in STEM and DFT simulations**. **a**, High-angle annular bright-field (HAABF) scanning transmission electron microscopy (STEM) plan view image of Cs-exchanged bilayer vermiculite. **b**, DFT simulation of image in (a). **c**, Cross-sectional HAABF-STEM image of natural vermiculite. The exchangeable ion is Mg. **d**, DFT simulation of image in (c). **e**, Cross-sectional HAABF-STEM image of Cs-exchanged vermiculite. **f**, DFT simulation of image in (e). All scale bars, 0.5 nm.

STEM imaging was performed using a probe-side aberration-corrected FEI Titan G2 80-200 kV operated at 200 kV. Images were collected using a convergence angle of 21 mrad and a high angle annular dark-field (HAADF) detector with an inner (outer) collection angle of 48(196) mrad, and a probe current of 12 pA. The Cs ions in the devices arranged in the crystalline pattern that agrees with our DFT simulations (Supplementary Fig. 3).

**Absence of lattice defects in proton-exchanged micas**

Micas are chemically very robust, and all previous studies have found that the proton-exchange process does not damage micas' aluminosilicate layers[12–20]. In our case, no damage was also expected, especially given the mild solution used for proton exchange (acetic acid). Nevertheless, we corroborated the stability of our micas by intentionally trying to dissolve them. If our acetic acid solution was heated to 80 °C, the mica crystals could not be dissolved, regardless of how many days the mica was left inside the solution. We only succeeded in dissolving micas by using an aggressive 1 M nitric acid solution at ~300 °C under reflux conditions for over one day. This result rules out any significant damage to our proton-exchanged mica devices. To further investigate if our mild acetic acid treatment could introduce any microscopic defects in the studied mica membranes, the proton-exchanged membranes were imaged using AFM and TEM. Supplementary Fig. 4a,b show that no defects could be detected with the AFM. The membranes were also inspected with TEM. This allowed atomic-scale resolution images over areas of several nanometers in size (note that Fig. 4c is a TEM rather STEM micrograph). Even with this resolution, no pinholes could be detected. These results are consistent with our mass spectrometry measurements, which ruled out defects of dimensions comparable to hydrogen molecules.

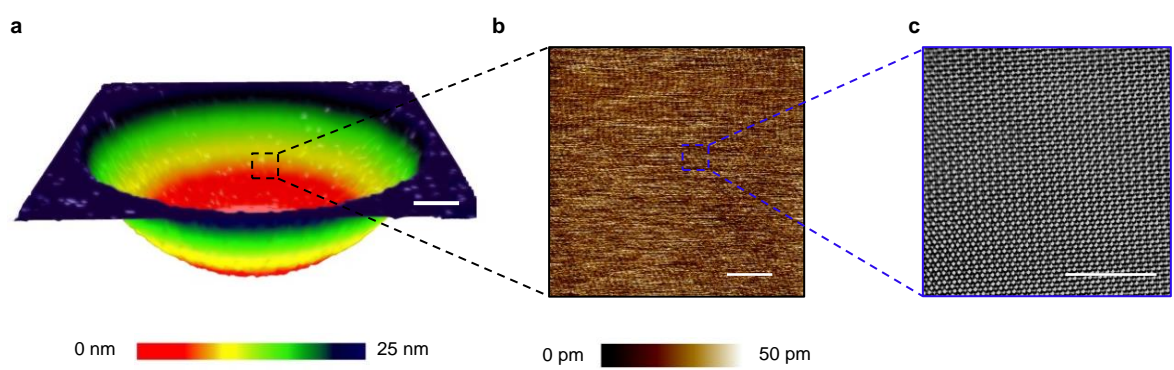

**Supplementary Figure 4 | Absence of lattice defects in proton-exchanged micas**. **a**, Atomic force microscopy image of a suspended proton-exchanged mica device. Scale bar, 500 nm **b**, AFM image zoomed in from the area marked in (a). Scale bar, 100 nm. **c**, Transmission electron microscopy image of the area in (b). Scale bar, 5 nm.

**Dependence of proton transport on humidity and hydrogen pressure**

The dependence of proton conductance on hydrogen content and humidity was studied using porous-Pt coated devices. Before discussing the experimental results, it is important to note that the Pt films in these devices are effectively infinite reservoirs of protons. First, Pt films can store large amounts of hydrogen, about one H atom per Pt atom[21]. Second, the Pt films cover an area ~$10^7$ times larger than the studied mica membranes (~3 µm$^2$). The latter suggests that, even at very low $H_2$ pressures, the Pt films would inject enough protons to saturate the mica, acting as infinite proton reservoirs. A similar argument applies for humidity. Despite the expectations, we investigated the dependence of conductance on both hydrogen pressure and humidity.

Let us discuss the humidity dependence first. Pt-coated devices were placed in a chamber with either dry (~10% humidity) or humid (100% humidity) $H_2$ atmosphere. Supplementary Fig. 5a shows that in humid hydrogen, both $G_x$ and $G_n$ were about 6 times larger than in the dry atmosphere. This increase could be reversed by changing the atmosphere back to dry. Next, the effect of hydrogen pressure was investigated. The devices were placed in a chamber at dry conditions and the $H_2$ partial pressure was gradually changed. Simultaneously, the proton current was measured as a function of time at a fixed voltage bias. Supplementary Fig. 5b shows that the current across the sample was clearly correlated with the hydrogen pressure. When the hydrogen pressure in the chamber increased from 1 mbar to 1 bar, the current across the sample increased by a factor of about 3. This increase could be reversed by reducing the hydrogen partial pressure to its original low value. These experiments demonstrate that proton conductance through our micrometer-size micas depended on both $H_2$ pressure and humidity.

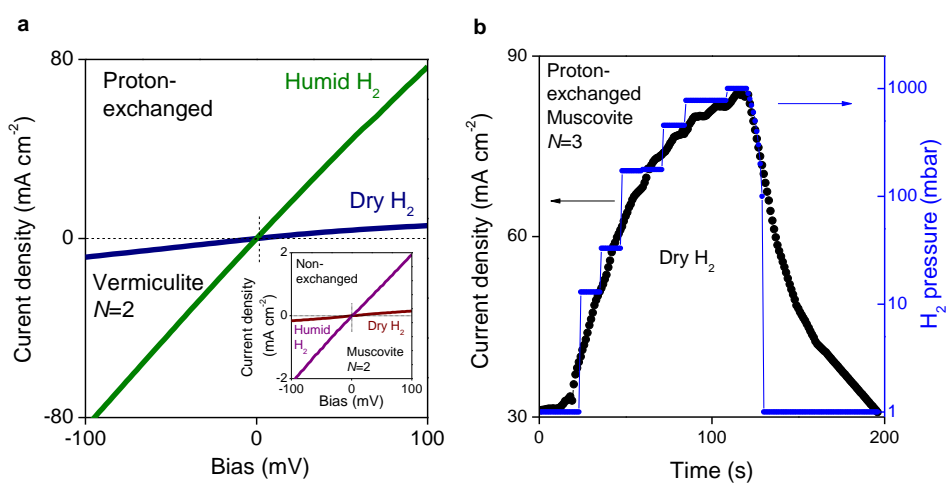

**Supplementary Figure 5 | Dependence of proton transport on humidity and hydrogen pressure**. **a**, *I-V* characteristics of a proton-exchanged mica in humid hydrogen (green) and dry hydrogen (blue) atmospheres. Inset: *I-V* characteristics of non-exchanged mica in humid hydrogen (purple) and dry hydrogen (brown) atmospheres. **b**, Left y-axis: current density recorded as a function of time for a proton-exchanged mica (black symbols). Applied bias, 1 V. Right y-axis: $H_2$ partial pressure inside the measurement chamber as a function of time (blue). Both data sets were recorded simultaneously.

**Reproducibility of Arrhenius plots**

Supplementary Fig. 6 shows that the Arrhenius plots $G_x(T)$ were highly reproducible both during heating and cooling cycles and for different devices. The activation energies in $G_x(T)$ were also reproducible for different devices. Below $T_p$, we found $E_p$ = 0.2±0.05 eV and, above $T_p$, $E_e$ = 0.83±0.06 eV. The experimental error in these measurements was similar to that in proton-transport measurements made using other 2D crystals[10]. The transition temperature itself was also reasonably reproducible for different devices (see Supplementary Fig. 6b) with $T_p$ = 215±35°C for muscovite and 170±10°C for vermiculite.

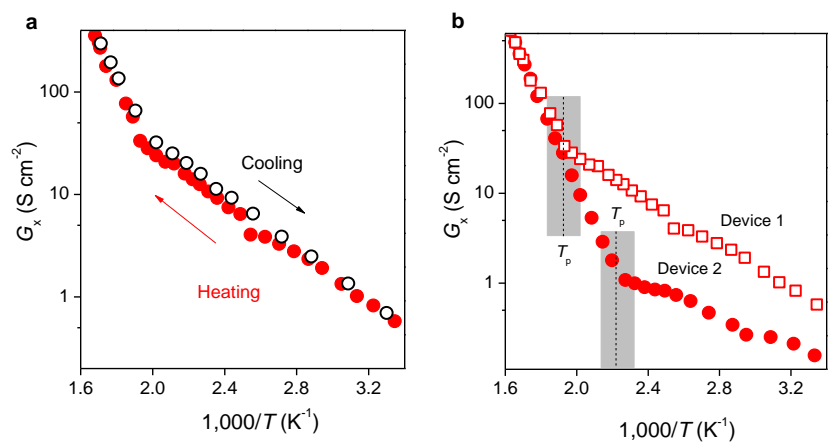

**Supplementary Figure 6 | Reproducibility of Arrhenius plots**. **a**, Arrhenius plots of $G_x$ from a bilayer muscovite device (Pt contact) during heating (red) and cooling (black) cycles. **b**, Arrhenius plots of $G_x$ from two bilayer muscovite devices (Pt contact). The transition temperature ($T_p$) for both devices is highlighted in grey.

**Density functional theory calculations**

Proton transport through micas was studied using density functional theory (DFT). The calculations were performed using the projector augmented wave (PAW) method, as implemented in the Vienna *ab-initio* Simulation Package[22–24] (VASP). To describe the electron exchange and correlation, the Perdew-Burke-Ernzerhof form of the generalized gradient approximation was adopted[25]. The van der Waals force, important for the layered materials, was taken into account by using the DFT-D2 method of Grimme[26]. The calculations were performed using the following parameters. The kinetic energy cutoff of the plane-wave basis set was at 500 eV in all calculations. The total energy difference between the sequential steps in the iterations was taken $10^{-5}$ eV as our convergence criterion. The convergence for the Hellmann-Feynman forces per unit cell was taken to be $10^{-4}$ eV Å$^{-1}$. Gaussian smearing of 0.05 eV was used.

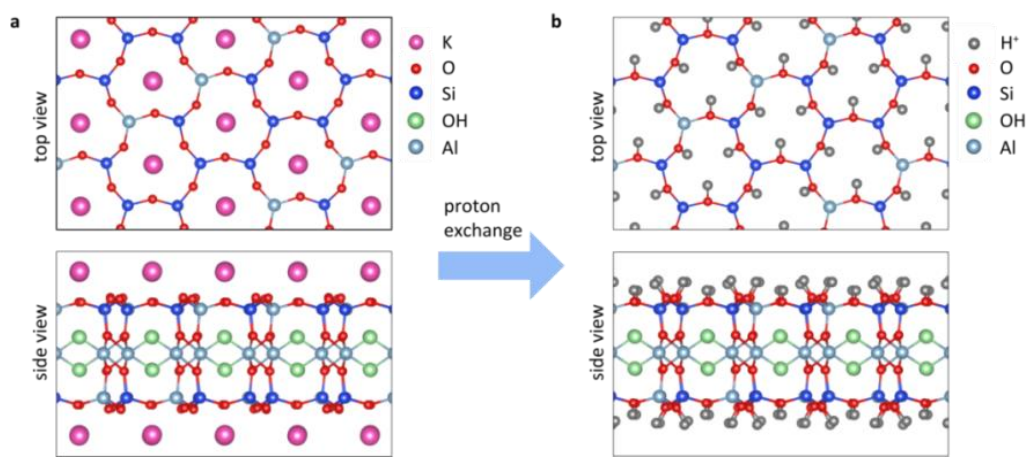

**Supplementary Figure 7 | Schematic illustration of proton exchange in muscovite layer**. **a**, Top and side views of non-exchanged muscovite. **b**, Top and side views of proton-exchanged muscovite layer. Each K ion is substituted by 3 protons.

We started by studying the proton exchange process itself. To that end, the energy of a monolayer muscovite crystal was calculated; first with the native ions (assuming $K^+$ in this case) and then with those K ions exchanged by protons (Supplementary Fig. 7a). We found that each K ion was substituted by 3 protons. Unlike $K^+$ that bonds to the center of the hexagonal ring in the basal plane, we found that protons bond to all the six oxygen atoms of the hexagonal ring, three for each exchanged K ion (see Supplementary Fig. 7b). The change in energy due to the proton exchange process was calculated using the formula: $\Delta E = E_s + nE_{H^+} - (E_{ex} + E_K)$ where $E_s$, $E_{ex}$, $E_{H^+}$, and $E_K$ are the total energies of the mica slab, the ion-exchanged slab, a single proton, and a single K ion, respectively. This calculation revealed that the proton exchange process was energetically favorable, leading to a gain $\Delta E$=0.92 eV per K ion.

Next, we calculated the minimum energy route for a proton as it transferred through proton-exchanged mica. To that end, a proton was moved in the perpendicular direction ($z$) to the mica basal plane in steps of ~0.5 Å. Near the minima, we decreased the step further down to ~0.25 Å. At each step, the proton was fixed only in the $z$-position but allowed to relax freely in the other two directions. This procedure ensured that the minimum energy trajectory was found. Supplementary Fig. 8a shows the proton trajectory (black dots) as found from such calculations. The proton initially moves along the center of the hexagonal ring. Later it moves towards the first of the two hydroxyl ($OH^-$) groups (green balls) inside the aluminosilicate layer. The proton is then transferred from one hydroxyl group to the next and finally exits through the hexagonal ring on the opposite side of the mica. Supplementary Fig. 8b shows the energy profile associated with this trajectory.

The energy profile is attractive and consists of six local minima. The local minima at both surfaces of the basal plane were found to be physisorption-like with a proton adsorption energy of ~0.37 eV. The second set of minima corresponds to a proton that resides at the interstitial void between the sublayers of Si and Al atoms and is bonded to the hydroxyl group. The corresponding adsorption energy is ~0.51 eV. The lowest energy minima are reached as the proton bonds to the oxygen atoms at the same sub-plane as the hydroxyl group. It should be pointed out that the energy difference between the lowest and the second lowest configurations is only ~0.06 eV. The existence of local minima inside the mica supports the hypothesis that protons jump along local traps as they transfer along the tubular channels. The calculations also allowed us to determine the activation energy barrier as the largest energy step along the trajectory. This approach leads to $E$~0.35 eV, which is in reasonable agreement with the experimentally found $E_p$≈0.2 eV. Even better agreement with the experiment can be expected if we take into account that the initial energy of incoming protons is lifted by zero-point oscillations at oxygen or Pt bonds[11,27].

Having studied proton transport through the mica lattice itself, let us also mention the role of adsorbed water on the surfaces of the aluminosilicate slabs. Water molecules are sterically excluded from the tubular channels of the mica lattice[13,28]. However, atomic force microscopy of water absorbed on proton-exchanged micas showed[29] that protons belonging to surface water molecules are attracted towards the OH groups inside micas' hexagonal rings (green balls in Supplementary Fig. 8). Therefore, protons of surface water slightly penetrate into the tubular channels trying to reach towards the OH groups inside mica[29]. In combination with our study, this result suggests that the first step in the proton transport mechanism for micas should be a proton ($H^+$) jump from surface $H_3O^+$ towards the $OH^-$ groups inside.

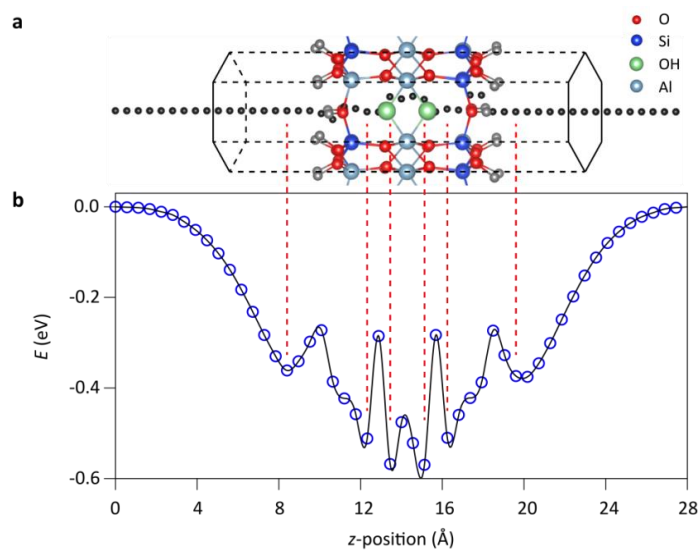

**Supplementary Figure 8 | DFT calculation of proton transport through micas**. **a**, Minimum energy path for H$^+$ transport through the basal plane of proton-exchanged muscovite. Localized H$^+$ after the proton-exchange process are shown by the grey balls to distinguish them from the translocating proton (black dots). **b**, Energy as function of the position of H$^+$. Local minima in the energy curve are found when H$^+$ is at the vicinity of hydroxyl groups and oxygen atoms. The red-dotted lines indicate positions of the translocating proton at the energy minima.

Having established the main ingredients of the mechanism of proton permeation through few-layer micas, it would be of interest in the future to consider the role of second-order structural features of aluminosilicate layers. For example, in trioctahedral micas (such as the studied vermiculite) the OH groups are oriented perpendicular to the basal plane whereas in dioctahedral micas (muscovite) they are leaning out of the plane[30]. This orientation difference was shown to affect the exchange rates of interlayer ions[31] and, therefore, it is possible that this also has repercussions for proton transport (for example, by shortening the distance between proton jumps in vermiculite).

**Potential for scaling up**

Similar to the industrial techniques developed for graphene, we envisage two potential routes for scaling up mica-based proton membranes. The first one is growth of ultrathin aluminosilicate films on metal or other substrates. For graphene, this approach was shown to produce monolayer films on a true industrial scale[32]. Encouragingly, initial reports show that a similar approach can be implemented for micas[33]. Such few-layer films grown initially on a substrate can then be transferred onto a porous support to form mechanically robust, gas impermeable membranes – similar to those already demonstrated for graphene[32,34,35]. The second route is to produce mica laminates via liquid exfoliation of bulk clay crystals. This approach is capable of producing exceptionally thin laminates (below 10 nm)[36]. Liquid exfoliation of clay minerals has already been reported[37] and with further effort[36], it might be possible to fabricate ultrathin mica laminates that preserve the gas impermeability and proton conducting properties reported in our work.